\documentstyle[graphicx,multicol,prb,aps]{revtex}
\input psfig.sty

\begin{document}
\title{A simplified method for the computation of correlation effects on the
band structure of semiconductors} 
\author{U. Birkenheuer$^1$, P. Fulde$^1$ and H. Stoll$^2$}
\address{$^1$Max-Planck-Institut f\"ur Physik komplexer Systeme, N\"othnitzer\\
Stra\ss e 38, 01187 Dresden, Germany\\
$^2$Universit\"at Stuttgart, Pfaffenwaldring 57, 70550 Stuttgart, Germany}
\date{\today}
\maketitle

\begin{abstract}
We present a simplified computational scheme in order to calculate the effects
of electron correlations on the energy bands of diamond and silicon. By
adopting a quasiparticle picture we compute first the relaxation and
polarization effects around an electron set into a conduction band Wannier
orbital. This is done by allowing the valence orbitals to relax within a
self-consistent field (SCF) calculation. The diagonal matrix element of the
Hamiltonian leads to a shift of the center of gravity of the conduction band
while the off-diagonal matrix elements result in a small reduction of the
conduction-electron band width. This calculation is supplemented by the
computation of the loss of ground state correlations due to the blocked Wannier
orbital into which the added electron has been placed. The same procedure
applies to the removal of an electron, i.e., to the valence bands. But the
latter have been calculated previously in some detail and previous results are
used in order to estimate the energy gap in the two materials.
The numerical data reported here shows that the methods works, in principle,
but that also some extension of the scheme is necessary to obtain fully
satisfactory results.
\end{abstract}

\vspace{1cm}

Dedicated to J.-P. Malrieu on the occasion of his 60th birthday.

\vspace{.5cm}

\begin{multicols}{2}
\section{Introduction}

Reliable energy-band calculations of solids remain a major topic of modern
condensed-matter theory. The problem is a very old one and has been treated on
various levels of accuracy and sophistication. Originally the calculations were
based on a one-electron theory, i.e., many-body effects were simply
neglected or indirectly accounted for~\cite{Harrison}. A good choice of model
potentials was crucial for the determination of energy bands, e.g., of
semiconductors or metals. After the development of density functional theory
(DFT) the field obtained an immense impact. Although the theory is designed for
ground-state properties it has been also used for the determination of energy
bands. Within the local density approximation (LDA) density functional theory
could be cast into the form of the Kohn-Sham equation~\cite{KohSha} containing
a self-consistent potential which is free of model-dependent assumptions. The
eigenvalues of that equation are simply interpreted as representing the energy
bands of the system. The computational efforts were significantly reduced and
simplified by the development of linearized methods~\cite{Andersen}. The
results agreed often surprisingly well with experiments indicating partial
error cancellations. In cases where the agreement is poor it is difficult to
make improvements which are free of arbitrariness. For example, for strongly
correlated electrons the LDA+$U$ method has been applied~\cite{AnZaAn} where
the intra-atomic correlations are described by a Hubbard interaction term with
an on-site Coulomb integral $U$. Thereby the problem of double counting of the
interactions must be carefully considered. More recently the LDA+$U$ has been
combined with the dynamical mean-field theory (DMFT)~\cite{AnPoKo}, an
extension of the coherent potential approximation (CPA) to include the effects
of dynamical fluctuations (dynamical CPA)~\cite{Kakehashi}, or the so-called
GW method~\cite{Hedin,StMaHa}. 

One of the shortcomings of the LDA when applied to semiconductors or insulators
is the underestimation of the size of the gap between the valence and
conduction bands. It is of no surprise that the LDA fails here because the
correlation cloud around an electron set into the conduction band differs
strongly from that of electrons in the charge-neutral ground state. While the
former includes a long-ranged polarization cloud the latter consists of van der
Waals fluctuations. But those two different physical situations cannot be
described by a single self-consistent potential depending on the density only
as assumed in the LDA. Therefore one would like to apply quantum-chemical
methods in order to calculate energy bands of semiconductors. They allow for
controlled approximations and can be done at different levels of accuracy.

Calculations based on quantum-chemical methods require a self-consistent field
(SCF) or Hartree-Fock (HF) energy-band calculation as a starting point. With
the CRYSTAL program package~\cite{CRYSTAL03} such calculations can be
done. Furthermore, because of the local character of the correlation hole
around an electron one also needs to know the Wannier functions for a given
semiconductor. Correlation effects arise from exciting electrons out of those
localized states. Wannier functions can be obtained either from the CRYSTAL
program package \cite{CRYSTAL03}
or from an alternative code called WANNIER~\cite{ShDoStFu}. While
it is straight forward to determine Wannier orbitals for valence bands it is
much more difficult to do the same for conduction bands, due to their weaker
localization and the frequently encountered mixing of the low-lying conduction
bands with energetically higher-lying bands.
But with a proper band disentanglement procedure this can be achieved
too~\cite{BirkIzotov}. It would be preferable to excite electrons out of
local non-orthogonal orbitals~\cite{StoFul} implying the use of a different set
of orbitals for a SCF and a correlation calculation. But in that case one
cannot use standard quantum chemical programs but rather has to develop
new ones.

In order to calculate energy bands for semiconductors with quantum chemical
methods one can proceed along two different routes. One starts from the
quasiparticle picture~\cite{HoHoFu,Graef97}. The real part of
the excitation energy is assumed to be much larger than the imaginary part so
that the life time of the excitation may be neglected. In that case one
identifies the most important microscopic processes which generate the
correlation hole around an electron or hole added to the system and
diagonalizes the Hamiltonian in that particular subspace. The latter includes
the additional electron or hole (projection method). A related method is based
on an effective Hamiltonian and has been exploited by Malrieu and
coworkers~\cite{AlReMa,Malrieu} and others~\cite{SunBar}. The second route
consists of calculating Green's
functions~\cite{LinOhr,Ladik,GrRuHe,AlbFul,Buth05},
the poles of which determine the excitation spectrum. In both cases an
incremental scheme~\cite{Stoll} can be applied when the correlation hole is 
constructed around the added particle or hole. A number of high-quality
calculations have been performed for the valence energy bands, e.g., of diamond
and silicon~\cite{Graef97,Graef93,Alb00}.
However, one is still not able to calculate the conduction bands
(and thus the energy gap) of these materials in the same way,
even with our new embedding tool~\cite{Birk-Emb,CRYMOLinter}
Convergence of the approach to an accuracy better than 0.5 eV seems presently
hard to achieve.

Here we want to suggest an alternative scheme for energy
bands which is based on the quasiparticle approach. It uses a division of the
correlation effects caused by the added 
particle (i.e., electron or hole) into two parts. One concerns polarization and
relaxation around the particle and another the loss of ground-state
correlations associated with its presence. A splitting of this form was studied
previously for a simplified model which allowed for a nearly analytic treatment
of those correlation contributions to the energy bands \cite{BorFul}. Here we
want to demonstrate that it can be used also for more accurate results based on
an {\it ab initio} Hamiltonian for a respectable basis set. Calculations of
this kind
seem to be much easier than previously used computational schemes for dealing
with the Hamiltonian.

The numerical results which we shall present here require further work and
improvements. That is related to the rather diffuse form of the Wannier functions
for conduction electrons in the covalent systems diamond and silicon. But it is
also demonstrated that the improvements are straight forward. 

\section{Description of the method}

We assume that the Hamiltonian is written for a basis set consisting of
Gauss-type orbitals (GTO's). Furthermore, we assume that the Hartree-Fock (HF)
or self-consistent field ground-state wave function $| \Phi_{\rm SCF} \rangle$
of the $N$-electron system has been determined. For the elemental
semiconductors we are interested in, this can be achieved by using the program
package CRYSTAL \cite{CRYSTAL03}. When an electron with momentum ${\bf k}$ and 
spin $\sigma$ is added to
the energy band $\nu$ the ($N$+1)-particle state is
\begin{equation}
  \mid \Phi^{N+1}_{\nu \sigma} ({\bf k}) \rangle = c_{{\bf k} \nu \sigma}^+
  \mid \Phi_{\rm SCF} \rangle~~~. 
\label{phiN+1}
\end{equation}
\noindent Similarly, when an electron is removed from the system the state is
\begin{equation}
  \mid \Phi^{N-1}_{\nu \sigma} ({\bf k}) \rangle = c_{{\bf k} \nu \sigma} 
  \mid \Phi_{\rm SCF} \rangle~~~. 
\label{phiN-1}
\end{equation}
Next we want to express the Bloch states as superposition of Wannier states.
From the occupied HF orbitals one can construct Wannier orbitals
without problems, e.g., by using the Wannier-Boys localization method 
\cite{Zicovich} implemented in CRYSTAL \cite{CRYSTAL03}. For the 
virtual space this is more difficult. The required procedure is described in
Ref.~\cite{BirkIzotov} to 
which we refer. In the following we assume that those orbitals have been
determined and that the corresponding creation (destruction) operator are
$c^+_{{\bf R} n \sigma}$ ($c_{{\bf R} n \sigma}$). Here ${\bf R}$ denotes the
lattice vector of the unit cell at which the Wannier orbital is centered while
$n$ is an intracell index. One can imagine the Wannier orbitals as representing
the bonding and antibonding orbitals in the diamond structure of the elemental
semiconductors. We write
\begin{equation}
  \mid \Phi^{N+1}_{\nu \sigma} ({\bf k}) \rangle = \frac{1}{\sqrt{N_c}}
  \sum_{{\bf R} n} \alpha_{\nu n} ({\bf k}) e^{i {\bf kR}} c_{{\bf R} n
  \sigma}^+ \mid \Phi_{\rm SCF} \rangle 
\label{nusigmak}
\end{equation}
\noindent and similarly for $|\Phi^{N-1}_{\nu \sigma} ({\bf k}) \rangle$. Here
$N_c$ is the number of unit cells. The conduction-band energies in Hartree-Fock
approximation can be expressed in terms of local matrix elements as
\begin{equation}
  \epsilon ^{\rm SCF}_{{\bf k} \nu} = \sum_{\bf R} \sum_{nn'} \alpha_{\nu n}
  ({\bf k}) \alpha^\ast_{\nu n'} ({\bf k}) e^{i {\bf kR}} H^{\rm SCF}_{{\bf R}
  nn'}
\label{epsSCF}
\end{equation}
\noindent with matrix elements
\begin{eqnarray}
  H^{\rm SCF}_{{\bf R} - {\bf R}', nn'} & = & \langle \Phi_{\rm SCF} \mid
  c_{{\bf R}' n' \sigma} Hc^+_{{\bf R} n \sigma} \mid \Phi_{\rm SCF} \rangle
  \nonumber \\ 
&& -\delta_{{\bf RR}'} \delta_{nn'} E^{\rm SCF}_0~~~.
\label{HscfR}
\end{eqnarray}
Here $ E^{\rm SCF}_0$ is the Hartree-Fock ground-state energy. When the
electronic 
correlations are included, the Hartree-Fock excitations become quasiparticle
excitations. For relatively weakly correlated electron systems one may assume
that there is still a one-to-one correspondence of the Hartree-Fock and
quasiparticle excitations. In this case the energy bands describe the
dispersion of the quasiparticles. Life-time effects are neglected. Thus we
write for the correlated ($N$+1)-electron state
\begin{eqnarray}
  \mid \Psi^{N+1}_{\nu \sigma} ({\bf k}) \rangle
  &=& \tilde{\Omega} \mid \Phi^{N+1}_{\nu \sigma} ({\bf k}) \rangle \nonumber \\
  &=& \frac{1}{\sqrt{N_c}}
  \sum_{{\bf R} n} \alpha_{\nu n} ({\bf k}) e^{i {\bf kR}} \: \tilde{\Omega}
  c_{{\bf R} n \sigma}^+ \mid \Phi_{\rm SCF} \rangle
\label{phinuSig}
\end{eqnarray}
\noindent where the wave- or M\"oller operator $\tilde{\Omega}$ transforms
$|\Phi^{N+1}_{\nu \sigma} ({\bf k}) \rangle$ into $|\Psi^{N+1}_{\nu \sigma}
({\bf k}) \rangle$. The correlated conduction bands are given by
\begin{equation}
  \epsilon_{\nu \sigma} ({\bf k}) = \frac{\langle \Phi^{N+1}_{\nu \sigma}
  ({\bf k}) \mid {\tilde{\Omega}}^+ H \tilde{\Omega} \mid \Phi^{N+1}_{\nu
  \sigma} ({\bf k}) \rangle}{\langle \Phi^{N+1}_{\nu \sigma} ({\bf k}) \mid
  {\tilde{\Omega}}^+ \tilde{\Omega} \mid \Phi^{N+1}_{\nu \sigma} ({\bf k})
  \rangle} - E_0
\label{epsnusigm}
\end{equation}
\noindent where $E_0$ is the exact, i.e., correlated ground-state energy. In
terms of cumulants this can be rewritten as
\begin{equation}
  \epsilon_{\nu \sigma} ({\bf k}) = (H \mid \Omega)_{{\bf k} \nu \sigma} - E_0
\label{epsk}
\end{equation}
\noindent where the notation
\begin{equation}
  (A \mid B)_{{\bf k} \nu\sigma} = \langle \Phi^{N+1}_{\nu \sigma} ({\bf k})
  \mid A^+ B \mid \Phi^{N+1}_{\nu \sigma} ({\bf k})\rangle^c
\label{ABk}
\end{equation}
\noindent has been used with the index $c$ indicating that the cumulant has to
be taken. For cumulants and their properties we refer to Ref.~\onlinecite{Fulde}
where also the original literature is cited. The
operator $\Omega$ is the cumulant wave operator, i.e., it is defined only in
connection with the form (\ref{ABk}). It is customary to write $\Omega=1+S$
where $S$ is a scattering matrix. By decomposing the Hamiltonian
$H=H_{\rm SCF}+H_{\rm res}$ we obtain for the correlation correction of the
conduction bands
  \begin{eqnarray}
  \epsilon_{\nu \sigma} ({\bf k}) &=& \epsilon_{\nu \sigma}^{\rm SCF} ({\bf k})
  + \Delta \epsilon_{\nu \sigma} ({\bf k}) \nonumber \\ 
  \Delta \epsilon_{\nu \sigma} ({\bf k}) & = & \sum_{\bf R} \sum_{nn'}
  \alpha_{\nu n} ({\bf k}) \alpha_{\nu n'}^* ({\bf k}) e^{i {\bf kR}} \nonumber
  \\  
&& \left( c_{on'\sigma}^+ \mid H_{\rm res}~ S~ c_{{\bf R} n \sigma} \right)~~~.
\label{ensigma}
\end{eqnarray}
\noindent The round brackets $(A|B)$ refer to the cumulants
\begin{equation}
  (A \mid B) = \langle \Phi_{\rm SCF} \mid A^+B \mid \Phi_{\rm SCF} \rangle^c~~.
\label{ABphi}
\end{equation}
Note that the cumulant $(c_{{\bf 0}n'\sigma}^+\mid H_{\rm res}~S~c_{{\bf R} n
\sigma}^+)$ contains a term $E_{\rm corr}=E_0-E_0^{\rm SCF}$ in form of a
subtraction which describes the correlation energy of the ground state of the
$N$-electron system.
It changes when differently localized Wannier
functions are used. Yet, since Wannier functions provide a complete basis,
the shifts $\Delta \epsilon_{\nu \sigma}$ remain the same after summation of
all relevant contributions according to (\ref{ensigma}).

For systems like the elemental semiconductors it suffices to include in $S$
one- and two-particle excitations. Thereby, the annihilation operators
$c_{{\bf R}n \sigma}$ refer to the occupied Wannier orbitals of the
($N$+1)-electron system while the number of the local creation operators
$a^+_{{\bf R}\tau \sigma}$ depends on the size of the basis set. The latter
are, e.g., the ones for GTO's with
the GTO being centered at site (or bond) $\tau$ of the unit
cell ${\bf R}$. Note that due to the use of cumulants only connected operator
products contribute.
Thus the following general ansatz for the scattering matrix $S$ holds
\begin{eqnarray}
  S & = & \sum_{{\bf R}_1{\bf R}_2} \sum_{\tau n} \sum_{\sigma \sigma'}
  \eta^{\tau n}_{{\bf R}_1{\bf R}_2} a^+_{{\bf R}_1 \tau \sigma} c_{{\bf R}_2 n
  \sigma} \nonumber \\
  & + & \sum_{{\bf R}_1{\bf R}_2{\bf R}_3{\bf R}_4}\sum_{\tau \tau' n n'}
  \sum_{\sigma \sigma'} \eta^{\tau \tau' n n'}_{{\bf R}_1{\bf R}_2{\bf R}_3{\bf
  R}_4} \nonumber \\
& \cdot & a^+_{{\bf R}_1 \tau \sigma} a^+_{{\bf R}_2 \tau' \sigma'} c_{{\bf
  R}_3 n \sigma'} c_{{\bf R}_4 n' \sigma} ~~~.
\label{SsumR}
\end{eqnarray}
\noindent It is advantageous to decompose $S$ into a part which commutes with
the added electron $c^+_{{\bf R} n \sigma}$ and a part which does not, i.e.,
\begin{equation}
  S = S_\pi + S_\eta
\label{SSS}
\end{equation}
\noindent with 
\begin{equation}
  \left[ S_\pi, c_{{\bf R} n \sigma}^+ \right]_- \neq 0~~ {\rm and} ~~ \left[
  S_\eta, c_{{\bf R} n \sigma}^+ \right]_- = 0 ~~~.
\label{pieta}
\end{equation}
\noindent The operator $S_\pi$ describes the relaxation and polarization
around the electron generated by $c_{{\bf R}n \sigma}^+$ while $S_\eta$
describes the loss of ground-state correlations due to presence of the extra
electron. As pointed out before, the latter results from blocking the orbital
occupied by the added electron for excitations out of the ground state. Thus
\begin{eqnarray}
  \left( c^+_{{\bf 0} n' \sigma} \mid H_{\rm res}~ S~ c^+_{{\bf R} n \sigma}
  \right)
  & = & \langle \Phi_{\rm SCF} \mid c_{{\bf 0} n' \sigma} H_{\rm res} S_\pi
  c^+_{{\bf R} n \sigma} \mid \Phi_{\rm SCF} \rangle^c \nonumber \\
  & + & \langle \Phi_{\rm SCF} \mid  c_{{\bf 0} n' \sigma} H_{\rm res}
  c^+_{{\bf R} n \sigma} S_\eta \mid \Phi_{\rm SCF}\rangle^c . \nonumber \\
\label{consigH}
\end{eqnarray}
The first term, i.e., the polarization and relaxation energy is approximated by
a SCF calculation for the ($N$+1)-electron system with an
electron frozen in Wannier orbital $({\bf R}n \sigma)$. This can be done by
making use of the program package MOLPRO~\cite{MOLPRO}. The energy contribution
of the second term is computed by determining those excitations in a
calculation of the ground-state correlation energy which are blocked by the
creation $c^+_{{\bf R}n \sigma}$ of an electron. Their contribution to the
energy can be extracted from a ground-state calculation. The details of this
procedure are described in the following subsection. Needless to say that the
same prescription applies for calculating the valence band, i.e., when an
electron is removed from the $N$-particle ground state.

Although the scheme proposed here consists of several individual steps 
and may seem cumbersome from a technical point of view, the present approach is 
conceptionally much simpler than other {\it ab initio} methods 
for band structures~\cite{Graef97,AlbFul,Buth05,Graef93,Alb00,Bez04}.
It is designed such that it only requires Hartree-Fock type calculations and
some small configuration interaction (CI) correlation calculations to estimate
the correlation effects on the energy bands. Both type of calculations can be
performed routinely and without much computational effort by standard quantum
chemical program packages.

\section{Computations and Results}

In a series of papers \cite{Graef97,Graef93,Alb00}, we presented results for
the influence of correlation effects on the valence band structure of diamond
and silicon. These results were obtained using a local Hamiltonian
formalism in finite cluster calculations for the neutral $N$-particle state and
cationic ($N$-1)-particle hole states. Unfortunately, the structure of the
conduction band is not accessible this way, since an additional electron in
negatively charged ($N$+1)-particle states of finite C and Si clusters is not
bound. Therefore, only rough estimates of the band gaps in diamond and silicon
could be given in Ref.~\onlinecite{Alb00}, assuming that the energetic effect
of the polarization cloud around a conduction band electron would be
approximately the same as that around a valence-band hole.

In the meantime, we adapted our local Hamiltonian formalism to the
treatment of conduction bands, by extracting the localized conduction band
Wannier
functions (WF) from periodic Hartree-Fock (HF) calculations
\cite{Zicovich,CRYSTAL0X}. The same calculations are used to generate embedding
potentials which describe the influence of the infinite bulk surroundings on
finite cluster models \cite{Birk-Emb}. Results for the band structure of
diamond \cite{WillBirk} and hydrogen fluoride chains \cite{Birk-HF} with this
approach are under preparation in our laboratory.

As an alternative, we present here results for the diamond and silicon
conduction bands using the simplified method described in the previous section.
Thereby, instead of generating local matrix elements of the system's
Hamiltonian between localized multi-determinant conduction band states, we
start from the simple physical picture of a polarization cloud forming around
an added conduction-band electron and calculate the (static) polarization and
relaxation effects around this electron when it is placed in a frozen
conduction band WF. This yields localized quasi-particle states which we can
use as a basis for the final diagonalization of the Hamiltonian. This in turn
should lead to a 
realistic quantitative description of correlation effects on the conduction
bands. The advantage over the local Hamiltonian approach lies in the facts
that by freezing the conduction band WF, we do not encounter stability
problems, and that for the determination of the static (rather than dynamic)
polarization cloud self-consistent field (SCF) calculations (rather than
correlated calculations) are sufficient. It is clear, on the other hand, that
the computational simplification cannot be achieved without introducing
additional approximations: the difference between static and dynamic
relaxation/polarization effects is neglected and, more importantly, correlation
effects beyond relaxation/polarization are also neglected at this stage. The
most important of the latter effects is the loss of ground-state correlation
when introducing an electron in the conduction band, i.e., the exclusion effect due to the non-availability of excitations from occupied orbitals of the
neutral ($N$-electron) system into the conduction band Wannier spin orbital
occupied in the ($N$+1)-electron system. However, the leading term of this
exclusion effect on the diagonal matrix element can easily be recovered in a
small configuration-interaction (CI) calculation, restricted to excitations
from the occupied SCF orbitals of the closed-shell $N$-electron system into a
single localized conduction band WF, without abandoning the advantages of
computational simplicity.

Following the above strategy, we performed the following detailed
computational steps:
\begin{enumerate}
\item[1)]
We started from periodic HF calculations for bulk diamond and silicon. The 
CRYSTAL200x program \cite{CRYSTAL0X} was used for that purpose, a pre-cursor
of the most recent public version of the CRYSTAL program package 
\cite{CRYSTAL03} which already allowed for localization. The basis sets used
were of valence-triple-zeta quality, with a single $d$ polarization function,
optimized for diamond \cite{dunning,Birk-Bas} and silicon \cite{Bergner93}.
In the case of Si, the [1$s$2$s$2$p$] core was simulated by an effective core
potential \cite{Bergner93}. These calculations provide the HF band structure
of the conduction (and valence) bands.
\item[2)]
The Wannier-Boys algorithm \cite{Zicovich,Baranek} is used to optimize the
unitary band mixing matrix of the multi-band Wannier transformation. It is
applied separately to the valence band and conduction band Bloch functions to
generate localized occupied and virtual Wannier functions. In the case of Si,
the conduction band states had to be disentangled from those of the
higher-lying unoccupied bands before they could be localized \cite{BirkIzotov}.
The mean width $\sqrt{\langle({\bf r}-\langle {\bf r}\rangle)^2\rangle}$ of
the resulting virtual Wannier functions is found to be in the order of 
1.5-2 times the bond length in the elemental semiconductor.
\item[3)]
The data of steps 1 and 2 were transferred via our CRYSTAL-MOLPRO interface
\cite{CRYMOLinter} to the molecular program MOLPRO \cite{MOLPRO,MOLPROci} and
are used as input for generating embedded C$_{8(18)}$ and Si$_{8(18)}$ clusters
\cite{WillBirk}. Here, the central C-C or Si-Si bond and its six
nearest-neighbor bonds inside the C$_8$ (Si$_8$) kernel are treated as
active, while the basis functions from the surrounding shell of 18 C (Si) atoms
are only added to assist in the representation of the localized occupied and
virtual orbitals of the kernel.
These C$_{26}$ and Si$_{26}$ support clusters are sufficient to 
provide good representations of the localized Wannier functions inside the
kernel. Upon projection onto the support clusters the norms of the virtual
Wannier functions reduce by 1-2\% only.
For the generation of the local virtual basis functions 
a compact variant 
\cite{Birk-Emb} of the projected atomic orbital construction \cite{Pulay-PAO}
is used. The Hartree-Fock potential of all non-active occupied C-C/Si-Si
orbitals on the cluster and the surrounding solid (together with the nuclei
there) is incorporated in the embedding potential.
\item[4)]
The HF reference data for matrix elements between localized ($N$+1)-particle
states can now be generated according to 
\begin{equation}\label{Eq:Hcorr}
  \langle \Phi_i^{N+1}   | H | \Phi_j^{N+1}   \rangle - 
  \langle \Phi_{\rm SCF} | H | \Phi_{\rm SCF} \rangle \delta_{ij}
\end{equation}
where $\Phi_{\rm SCF}$ is the HF wave function of the neutral C$_{8(18)}$ or
Si$_{8(18)}$ cluster, and $\Phi_i^{N+1}$ is the same wave function with a
localized (anti-bonding) conduction band WF (from step 2) added at bond $i$.
The latter describes the local HF configurations
$c^+_{{\bf R}n\sigma} \Phi_{\rm SCF}$ introduced in Eq.~(\ref{nusigmak}).
\item[5)]
To estimate the correlation effects for the diagonal matrix element ($i=j$),
we first replace $\Phi_i^{N+1}$ in Eq.~(\ref{Eq:Hcorr}) by
$\tilde{\Phi}_i^{(N+1)}$ where the valence orbitals are allowed to relax and
polarize due to the presence of the (fixed) conduction band WF of the
($N$+1)-particle system. This single-determinant wave function serves as a
first approximation to the correlated counter parts
$\tilde\Omega c^+_{{\bf R}n\sigma} \Phi_{\rm SCF}$ from Eq.~(\ref{phinuSig}).
\item[6)]
To include the leading term to the loss of ground state correlation
for the diagonal matrix element, we further replace $\tilde{\Phi}_i^{N+1}$ by
$\tilde{\Psi}_i^{N+1}$, and $\Phi_{\rm SCF}$ by $\tilde{\Psi}^{N}$ in
Eq.~(\ref{Eq:Hcorr}), where $\tilde{\Psi}_i^{N+1}$ and $\tilde{\Psi}^{N}$
are full CI wave functions in the space of the active valence band orbitals
(from step 4 for the N-particle system, and step 5 for the (N+1)-particle
system) plus the localized conduction band WF at site $i$. (We also tried
including additional excitation channels with one electron in one of the
conduction-band WF $j$ adjacent to $i$, but the effect on the numerical results
to be described below was negligible [$<$0.05 eV].)
\item[7)]
In order to get the leading term for the correlation effects on the
off-diagonal matrix element between nearest-neighbor bond sites $i\neq j$, we
focus on the relaxation and polarization effect and approximately evaluate 
\begin{eqnarray}\label{2}
\langle \tilde{\Phi}_i^{N+1} | H | \tilde{\Phi}_j^{N+1} \rangle & \approx &
\langle \Phi_i^{N+1} | H | \Phi _j^{N+1} \rangle \nonumber \\
&& + 2 \langle \Phi _i^{N+1} | H | \tilde{\Phi}_j^{N+1}-\Phi_j^{N+1}\rangle
\end{eqnarray}
exploiting the fact that $\tilde{\Phi}_i^{N+1}$ and $\tilde{\Phi}_j^{N+1}$ are
equivalent in diamond lattices. To this end the polarized active valence 
orbitals $\tilde{\varphi}_n$ entering the determinant $\tilde{\Phi}_j^{N+1}$
are expanded into the unpolarized ones ($\varphi_n$) defining $\Phi_j^{N+1}$
plus orthogonal relaxation contributions $\omega_k$ in the virtual space
\begin{equation}\label{3}
\tilde{\varphi}_n = \sqrt{1-c^2_n}~ \varphi_n + c_n \omega_n \quad,
\end{equation}
and $\tilde{\Phi}_j^{N+1}$ is replaced by a CI wave function with single
excitations from the $\varphi_n$ into the $\omega_n$.
\end{enumerate}
\noindent The resulting SCF and correlation contributions are listed in Table 1.
The employed Wannier functions are rather compact, and the matrix elements of
the local Hamiltonian should decay rapidly with increasing distance of the bond
pairs, such that the diagonal and nearest-neighbor off-diagonal terms should 
suffice to recover most of the correlation effects. Switching to different
localized Wannier functions would alter these contributions slightly, but after
diagonalization of the resulting Hamiltonian matrix these modifications
cancel to a large extent. At present, we do not have any alternative
localization scheme for
Wannier functions at hand to corroborate this numerically.

The above diagonal correlation contributions lead to a shift of the center of
gravity of the band and sum up to --1.08 eV for diamond and --0.85 eV for
silicon. These energies are significantly smaller in magnitude than the
corresponding values found in Ref.~\onlinecite{Alb00} for the valence band
(relaxation/polarization: 4.6 eV for C, 3.4 eV for Si; loss of ground-state
correlation: --0.7 eV for C, --0.6 eV for Si). This finding is in strong
contrast to an assumption put forward in Ref.~\onlinecite{Alb00}, namely that
the correlation effects caused by conduction electrons and valence holes are
symmetric and presumably has to be attributed to the fact, that the
conduction band WFs are much less localized than the valence band WFs.
Still, the assumption is bound to hold for long-range polarization effects,
since the polarizing charges are of the same magnitude ($\pm e$), and if we
extract the polarization contributions beyond nearest-neighbor bonds (not
calculated in the present paper) from the valence-band calculations of
Ref.~\onlinecite{Alb00} (3.5 eV for C, and 2.4 eV for Si, without two-
and three-body contributions), we can estimate
the total correlation-induced shift of the conduction band center of gravity
to --4.6 eV for C, and --3.3 eV for Si.
A much simpler estimate of the long-range polarization contribution to 
the shift of the energy bands can be obtained from the classical 
polarization energy of a dielectric medium outside a sphere of radius
$R_{\rm cut}$ (see Ref.~\onlinecite{Fulde})
\begin{equation}\label{classEpol}
  \Delta \epsilon = - \frac{\epsilon_0 - 1}{2\epsilon_0}
                    \:\frac{e^2}{R_{\rm cut}}
  \quad.
\end{equation}
with $\epsilon_0$ being the dielectric constant of the medium. Using the experimental values for the dielectric constant
\vspace{1mm}

%%%%%%%%%%%%%%%%%%%%%%%%%%%%%%%%%%%%%%%%%%%%%%%%%%%%%%%%%%%%%%%%%%%%%%%%%%%%%%%
\begin{table}[htb]
\caption{Contributions to the diagonal and off-diagonal Hamiltonian matrix
elements (in eV) between localized conduction band states in bulk diamond
and silicon at various levels of theory (cf.\ text).}
\begin{tabular}{lrrrr}
  & \multicolumn{2}{c}{diagonal} & \multicolumn{2}{c}{off-diagonal} \\
  & diamond & silicon & diamond & silicon \\
\hline
  Hartree-Fock                       &  14.77 &   7.07 &   0.23 &   0.20 \\
  relaxation/polarization            & --1.28 & --0.98 & --0.01 & --0.02 \\
  loss of ground-state corr.         &   0.20 &   0.13 &        &        \\
  long-range rel./pol.$^a$           & --3.5  & --2.4  &        &        \\
  \hline
\multicolumn{5}{l}{$^a$ estimate based on the date
                        in Ref.~\onlinecite{Alb00}}\\
\end{tabular}
\end{table}
%%%%%%%%%%%%%%%%%%%%%%%%%%%%%%%%%%%%%%%%%%%%%%%%%%%%%%%%%%%%%%%%%%%%%%%%%%%%%%%

\noindent (5.7 for diamond and 12.0 for silicon)~\cite{expeps0}
the above estimates of the long-range polarization contributions correspond
to cutoff radii of 1.72~{\AA} for diamond and 2.75~{\AA} for silicon.
To gain some insight into the accuracy of the long-range 
estimates these cutoff radii can be compared to the radii of the 8 atomic
cluster kernels (defined here as the distance of the outermost kernel
atoms from the center of the kernel). They amount to 1.89 and 2.88~{\AA}
for C and Si, respectively, which is fairly close.

Turning now to the off-diagonal matrix elements between nearest-neighbor
bonds we find again here that the correlation-induced changes are much smaller
than for the corresponding valence band quantities evaluated in
Refs.~[\onlinecite{Graef97,Graef93}] (changes in the matrix elements of 0.7 eV
for C and 0.8 eV for Si). In diamond lattices the nearest-neighbor off-diagonal
elements enter with a factor of 8 into the bandwidth at the $\Gamma$-point
of the uppermost (lowermost) four valence (conduction) bands and with a factor
of 2 into the energy at the top (bottom) of the respective band complex.
It is essentially these contribution from the nearest-neighbor
matrix elements which is responsible for the total correlation-induced change
of the
valence-band width (of $\sim$7 eV for both, C and Si \cite{Graef97,Graef93}).
Assuming a similar finding for the conduction band, i.e., that only
matrix elements between nearest neighbors are important we
would estimate the  
correlation-induced narrowing of its width to be in the order of 0.1-0.2 eV.
The rising of its bottom would remain well below 0.05 eV, an almost negligible 
shift. Yet, this assumption is somewhat unrealistic, because the larger extent
of the conduction band WFs will lead to a much longer range of the
correlation-induced changes of the local matrix elements than in the
valence band case, i.e., to substantial changes far beyond nearest neighbours.
This is corroborated by our ongoing study of the conduction bands of diamond 
by means of the local Hamiltonian approach \cite{WillBirk}. Unfortunately, at
present we do not have estimated data for these changes available.

Indeed, our estimates for the center-of-gravity shifts (cf. above), when taken
together with the top-of-valence-band shifts of Ref.~\onlinecite{Alb00} (i.e.,
3.90 eV for C and 2.63 eV for Si) would lead to significantly too small band
gaps (5.3 eV vs.\ 7.3 eV from experiment \cite{expgaps} for C, and 2.5 eV
vs.\ 3.4 eV from experiment (direct gap) \cite{expgaps} for Si). They are
certainly more accurate than the (too large) HF band gaps of 13.8 eV for C,
and 8.4 eV for Si \cite{HFgaps}, but are still not acceptable. More accurate
data for the correlation-induced narrowing of the conduction band width are
needed, and also the determination of the polarization contributions beyond
nearest-neighbor bonds must be improved. Work in these directions is underway
in our laboratory.  
\vspace*{-3mm}

\section*{Acknowledgement}
We thank Dr. C. Willnauer for a number of discussions concerning numerical
results.  

\end{multicols}

\begin{references}
\bibitem{Birk-Bas} The most diffuse $s$ exponent was replaced by 0.201142,
the two outermost $p$ exponents by 0.625572 and 0.324363; the $d$ exponent
of 0.55 was taken from the corresponding double zeta basis.
\end{references}
\end{document}